\documentclass[preprintnumbers,floatfix,letterpaper,superscriptaddress,nofootinbib,twocolumn]{revtex4-2}
\usepackage{amsmath}
\usepackage{amssymb}
\usepackage{bbm}
\usepackage{subfigure}
\usepackage{hyperref}
\usepackage{url}
\usepackage[dvipdfmx]{graphicx}
\usepackage{xcolor}
\usepackage{color}
\usepackage{mathrsfs}
\usepackage{amsfonts}
\usepackage{latexsym}
\usepackage{ragged2e}
\usepackage{epsfig}
\usepackage{textcomp}
\usepackage{phaistos}
\usepackage[title]{appendix}
\usepackage{chngcntr}

\makeatletter
\renewcommand\@makefnmark{\hbox{\@textsuperscript{\normalfont\color{purple}\@thefnmark}}}
\renewcommand\@makefntext[1]{%
	\parindent 1em\noindent
	\hb@xt@1.8em{%
		\hss\@textsuperscript{\normalfont\@thefnmark}}#1}
\makeatother

\usepackage{caption}
\DeclareCaptionJustification{justified}{\leftskip=0pt \rightskip=0pt \parfillskip=0pt plus 1fil}
\graphicspath{{./Figures/}}

\def\beq{\begin{equation}}
\def\eeq{\end{equation}}

\def\mathbb{\Bbb}

\definecolor{vividviolet}{rgb}{0.62, 0.0, 1.0}
\definecolor{amaranth}{rgb}{0.9, 0.17, 0.31}
\definecolor{palatinateblue}{rgb}{0.15, 0.23, 0.89}
\definecolor{brightpink}{rgb}{1.0, 0.0, 0.5}
\definecolor{cornflowerblue}{rgb}{0.39, 0.58, 0.93}
\definecolor{deepcarminepink}{rgb}{0.94, 0.19, 0.22}
\definecolor{radicalred}{rgb}{1.0, 0.21, 0.37}
\colorlet{RED}{red}

\hypersetup{linktoc=all,
	colorlinks, linkcolor={palatinateblue},
	citecolor={brightpink}, urlcolor={amaranth}
}

\renewcommand{\d}[1]{\ensuremath{\operatorname{d}\!{#1}}}

\renewcommand{\d}[1]{\ensuremath{\operatorname{d}\!{#1}}}
\makeatletter
\def\@fnsymbol#1{\ensuremath{\ifcase#1\or $\textleaf$ \or $\PHplaneTree$
		\else\@ctrerr\fi}}%

\makeatother

\def\sideremark#1{\ifvmode\leavevmode\fi\vadjust{\vbox to0pt{\vss
			\hbox to 0pt{\hskip\hsize\hskip1em
				\vbox{\hsize1.5cm\tiny\raggedright\pretolerance10000
					\noindent #1\hfill}\hss}\vbox to8pt{\vfil}\vss}}}

\begin{document}
	\title{Feasibility of Primordial Black Hole Remnants as Dark Matter\\ in View of Hawking Radiation Recoil}
	
	\author{Sofia Di Gennaro}
	\affiliation{Center for Gravitation and Cosmology, College of Physical Science and Technology, Yangzhou University, \\180 Siwangting Road, Yangzhou City, Jiangsu Province  225002, China}

	\author{Yen Chin \surname{Ong}}
	\email{ycong@yzu.edu.cn}
	\affiliation{Center for Gravitation and Cosmology, College of Physical Science and Technology, Yangzhou University, \\180 Siwangting Road, Yangzhou City, Jiangsu Province  225002, China}
	\affiliation{School of Aeronautics and Astronautics, Shanghai Jiao Tong University, Shanghai 200240, China}

	\begin{abstract}
		It has recently been suggested that black hole remnants of primordial origin are not a viable dark matter candidate since they would have far too large a velocity due to the recoil of Hawking radiation. 
		We re-examined this interesting claim in more details and found that it does not rule out such a possibility. On the contrary, for models based on non-commutativity of spacetime near the Planck scale, essentially the same argument can be used to estimate the scale at which the non-commutativity effect becomes important. If dark matter ``particles'' are non-commutative black holes that have passed the maximum temperature, this implies that the non-commutative scale is about 100 times the Planck length. The same analysis applies to other black hole remnants whose temperature reaches a maximum before cooling off, for example, black holes in asymptotically safe gravity.
		\begin{center}
			
		\end{center}
	\end{abstract}
	
	\maketitle
	
	\section{Black Hole Remnants as Dark Matter Candidate}\label{intro}
	
	Dark matter, a mysterious component in our Universe that contributes additional gravitational pull and is essential for structure formation, remains elusive even after decades of efforts to pin down its identity. Theorists have proposed a plethora of dark matter candidates, ranging from various weakly interacting massive particles (WIMPs) to the attempts to do away with actual dark matter by modifying general relativity (which is a nontrivial task \cite{9410002}). Yet another possibility is that dark matter actually consists of microscopic primordial black holes. Of course, there could be many types of dark matter, so these possibilities are not mutually exclusive.
	
	Primordial black holes \cite{0801.0116} can be created from fluctuations in the early Universe \cite{Hawking, Hawking-Carr, GPY, Kapusta}. They may also be the result of phase transitions in the early Universe \cite{513978,9912422, 0706.1111}, or induced by inflation \cite{9404061,NORDITA-94-12-A,0406514,1709.05565,1712.09896,1712.09896, 1905.01741, 1907.04125, 1910.05238,1912.10437,2005.02895,2104.03972}.
	Indeed the possibility of primordial black holes playing the role of cold dark matter has been discussed extensively in the literature \cite{0205106, 0511743, 2006.02838, 2007.10722, 1906.11549, 2011.01930, 2103.04739}, and observational constraints from various approaches are now available \cite{2002.12778, 1906.09994, 1912.01014, 2004.00627, 2007.11804}. See also \cite{1510.01351,1812.11011,1902.08187,2001.09160,2003.10455,2102.06714} for future observational prospects.
	Notably, if dark matter consists of (entirely or mainly) primordial black holes, then the corresponding gravitational waves induced by non-Gaussian scalar perturbations are detectable by LISA-like detectors with some distinguishable features \cite{1810.11000}. The fraction of primordial black holes contributing to dark matter can also be constrained by measuring (anti)neutrino signals at the large liquid-scintillator detector of Jiangmen Underground Neutrino Observatory (JUNO) \cite{2010.16053}. Intriguingly, the NANOGrav results\footnote{In a co-dark matter scenario composed of both WIMPS and primordial black holes, it was found that constraints over solar-mass primordial black holes are consistent with the fraction in primordial black holes needed to explain NANOGrav results \cite{2011.05922}.} are consistent with possible signals from primordial black holes \cite{2009.07832, 2009.08268, 2010.03976}. Due to quantum gravitational corrections, it is possible that Hawking radiation either completely stops or slows down considerably for these black holes so they do not completely evaporate once they reach a certain size around the Planck scale. These are called black hole remnants \cite{0205106,1412.8366}. 
	
	Recently, Kov\'{a}\v{c}ik \cite{13} argued that microscopic black holes gain large velocities from the recoil of Hawking emissions and as such they would not be a feasible cold dark matter candidate (``cold'' means that their velocity is small). Indeed, as black holes evaporate due to Hawking radiation, they will recoil due to the conservation of linear momentum. Such an effect, accumulated over the typically long lifetime of an astrophysical black holes has been known to cause the position of the black holes to drift across space in the style of a random walk\footnote{Black hole recoil in the braneworld scenario was also investigated in \cite{0206046,0208102}.} \cite{page1980, 1210.6348}. The argument in \cite{13} is to consider some models in non-commutative spacetime in which the Schwarzschild-like black hole is essentially Schwarzschild at large enough mass, but its Hawking temperature eventually reaches a maximum, $T_m$, when the size of the black hole is sufficiently small. Afterwards the temperature continues to drop towards zero as the mass progressively decreases. Thus eventually the black hole becomes an extremal remnant with no further emission of particles. Crucially, the temperature as a function of the mass grows and falls considerably around the peak.
	
	Working in the units $G=c=\hbar=1$ (and sign convention $(-,+,+,+)$), we denote the mass of the black hole at its maximum temperature by $M_m$ and that of the final remnant mass by $M_0$. Due to the appreciable rise and drop of the temperature curve around the maximum,  the total number of quanta emitted as the black hole evaporates from mass $M_m$ to $M_0$ is relatively small \cite{13}: $N \sim (M_m-M_0)/T_m \lesssim 100$. Each quantum carries a momentum $p \sim (M_m-M_0)/N$, so after radiating $N$ particles the black hole receives a momentum recoil of $(M_m-M_0)/\sqrt{N}$ by the standard random walk argument. The final velocity of the black hole remnant is then 
	\begin{equation}
	v_r \approx \frac{M_m-M_0}{M_0 \sqrt{N}}.
	\end{equation}
	It was estimated in \cite{13} that the velocity is high $\left(\mathcal{O}(0.01)\right.$ to $\mathcal{O}(0.1)$ of the speed of light, depending on the models$\left. \right)$, which is far too large for these black holes to qualify as cold dark matter.
	
	In this work, we wish to argue that while the calculation is essentially correct, it does not rule out primordial black hole remnants as a possible viable dark matter candidate. There are a few reasons for this:
	Firstly, the non-commutative models come with the so-called non-commutative parameter, which dictates when non-commutativity effect becomes apparent. As will be discussed in Sec.(\ref{main}), the parameter is constrained by observations, and need not be taken to be equal to the Planck length as was done in \cite{13}. Indeed, we could turn this argument around and use the expected dark matter speed to deduce the non-commutative parameter. In fact, this is our main motivation -- to estimate the scale at which quantum gravitational effect might become important\footnote{Since we still lack a full quantum gravity theory, it is inevitable to discuss quantum gravity inspired phenomenological models, the downside of which is that most results are model dependent. So our analysis only applies to the non-commutative models and some other models with similar properties.}, by making contacts with observations. Secondly, the whole dynamics and time scale of the Hawking evaporation should be taken into account: if the initial mass of the black hole is large enough, it may not have enough time to evolve past the temperature peak. Furthermore, since the remnant has zero temperature, most likely the evaporation time scale from the temperature peak to zero temperature would take an infinite amount of time, which is to be expected from the third law of black hole thermodynamics. Lastly, this argument does not apply to models in which the black hole remnant stops evaporating at the maximum temperature, that is $M_m=M_0$, such as the GUP (generalized uncertainty principle)-corrected Schwarzschild black hole \cite{0106080}. 
	
	Let us now examine in details one of the non-commutative black hole models and comment more on the points mentioned above. More models are further explored in the Appendix.
	
	\section{Primordial Black Holes as Dark Matter in Non-Commutative Models}\label{main}
	
	As explained in \cite{1001.1205}, it is not possible to measure with arbitrary precision small volume of space arbitrary precisely. Any attempt to measure physical properties in a region of about the Planck size would involve so much energy that the local geometry should fluctuate significantly and therefore cause black holes (or other more exotic objects such as wormholes) to form. Intuitively one can understand this as a consequence of the uncertainty principle: if $\Delta x$ is small then $\Delta p$ of the probe (say, a light pulse) must be large. The energy $\Delta E = \Delta p \cdot c$ is therefore also large, and huge amount of energy fluctuation in a sufficiently small region would lead to black hole formation. 
	
	One possible mathematical model to implement such an idea geometrically is the so-called non-commutativity spacetime (see the review \cite{14a, 1240157, 0106048, 1203.6191}): at around the Planck scale, the spatial coordinates of the spacetime manifolds are subjected to the non-commutativity effect:
	\begin{equation}
	\left[\hat{x_i},\hat{x_j}\right] = 2i L \hat{x_k}\epsilon^{ijk},
	\end{equation}
	where $\epsilon^{ijk}$ is the Levi-Civita symbol and $L = \lambda \ell_\text{P}$ is the length scale of non-commutativity. 
	Non-commutativity is not just an \emph{ad hoc} proposal -- it is well motivated from more fundamental quantum gravity theories such as string theory \cite{9908142,0703173}.
	Since the effects of non-commutativity have not been observed in the experiments done so far, the parameter $L$ must satisfy the constraint $L < 10^{-18}$ m \cite{14}. See also \cite{1901.01613}, in which the parameter $L$ is argued to be around $10^{-21}-10^{-23}$ m by considering gamma ray burst (GRB) data (since non-commutativity can influence the group velocity of massless particle wave packets). The latter bound is only indicative of the upper bound due to uncertainties of the spectral lack and the GRB statistics.   
	
	Mathematically one could formally treat the mass of the Schwarzschild black hole in general relativity as being ``localized'' at the ``origin'' $r=0$, i.e. supported by a delta function. This is only a formality because the origin is spacelike and lies in the future of any observer. In the case of non-commutative spacetime, the ``source'' of the black hole is no longer sharply localized in spacetime by a delta function; rather, we need to use a matter density distribution that is centered in the origin. A possible distribution function is given by
		\begin{equation}
		\rho(r) = \rho_0 e^{-\left[r/(\alpha L)\right]^\alpha} \label{generic density},
		\end{equation}
	where $\alpha = 1,2,3$ is a parameter determining which model we are considering and $\rho_0$ also depends on the model.
	The density in \eqref{generic density} reproduces a delta function in the limit $\lambda\to 0 $.

	Let us start with the $\alpha = 2$ model described in \cite{14}.
	The models with $\alpha = 1$ and $\alpha = 3$ are similar and are shown in the appendix for completeness.
	From now on, we shall use natural units except where explicitly specified.
	
	The density in \eqref{generic density} in this case is given by
	\begin{equation}
	\rho(r) = \frac{M}{(2\lambda\sqrt{\pi})^3} e^{-\left[r/(2\lambda)\right]^2}.
	\end{equation}
	One considers a spherically symmetric Schwarzschild-like metric and imposes the conditions : $g_{00} = -g_{rr}^{-1}$ and $T_r^r = T^0_0 = -\rho(r)$ on the energy-momentum tensor. We can thus write the metric in the form: $g = \text{diag}\left(f(r), -f(r)^{-1}, r^2, r^2\sin(\theta)\right)$. The $00$-component of the Einstein equations thus gives an equation for $f(r)$:
	\begin{equation}
	\frac{1+f(r)+f^\prime(r)}{r^2} = 8\pi T_{00} = \frac{4M}{\lambda^3\sqrt{\pi}} e^{-\left[r/(2\lambda)\right]^2}.
	\end{equation}
	The solution of this equation is:
	\begin{equation}
	f(r) = -1 + \frac{4M}{\sqrt{\pi} r} \gamma\left(\frac{3}{2},\frac{r^2}{4\lambda^2}\right), \label{f(r)}
	\end{equation}
	where $\gamma\left(3/2,r^2/(4\lambda^2)\right)$ is the lower incomplete Gamma function defined as:
	\begin{equation}
	\gamma \left(\frac{3}{2},\frac{r^2}{4\lambda^2}\right) := \int_{0}^{\frac{r^2}{4\lambda^2}}  e^{-t} \sqrt{t} \d t.
	\end{equation}
	The event horizon is located at the root of $g_{rr} = 0$, i.e., at $f(r)=0$, corresponding to:
	\begin{equation}
	r_H = \frac{4M}{\sqrt{\pi}} \gamma\left(\frac{3}{2},\frac{r_H^2}{4\lambda^2}\right). \label{hor rad}
	\end{equation}
	This equation has to be solved numerically by plotting $f(r)$ and observing its intersections with the $r$ axis.
	\begin{figure}[h!]
		\centering
		\includegraphics[scale=0.60]{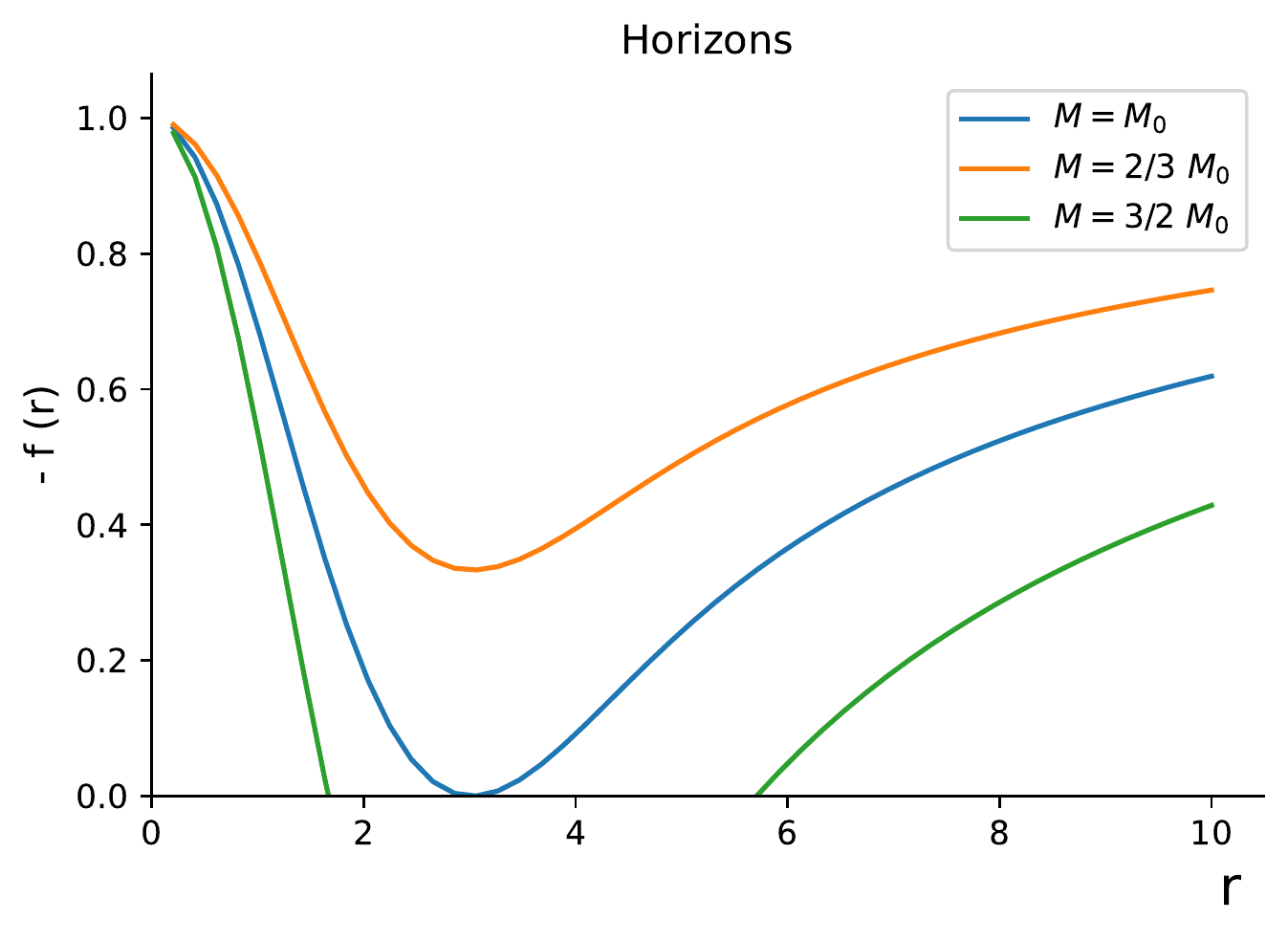}
		\caption{$-f(r)$ as a function of the coordinate radius, in units of $\lambda$ for $M=M_0$, $M =2/3$ $M_0$, and $M = 3/2$ $ M_0$.}
		\label{fig:horizons}
	\end{figure}
	
\begin{figure}[h!]
		\centering
		\includegraphics[scale=0.60]{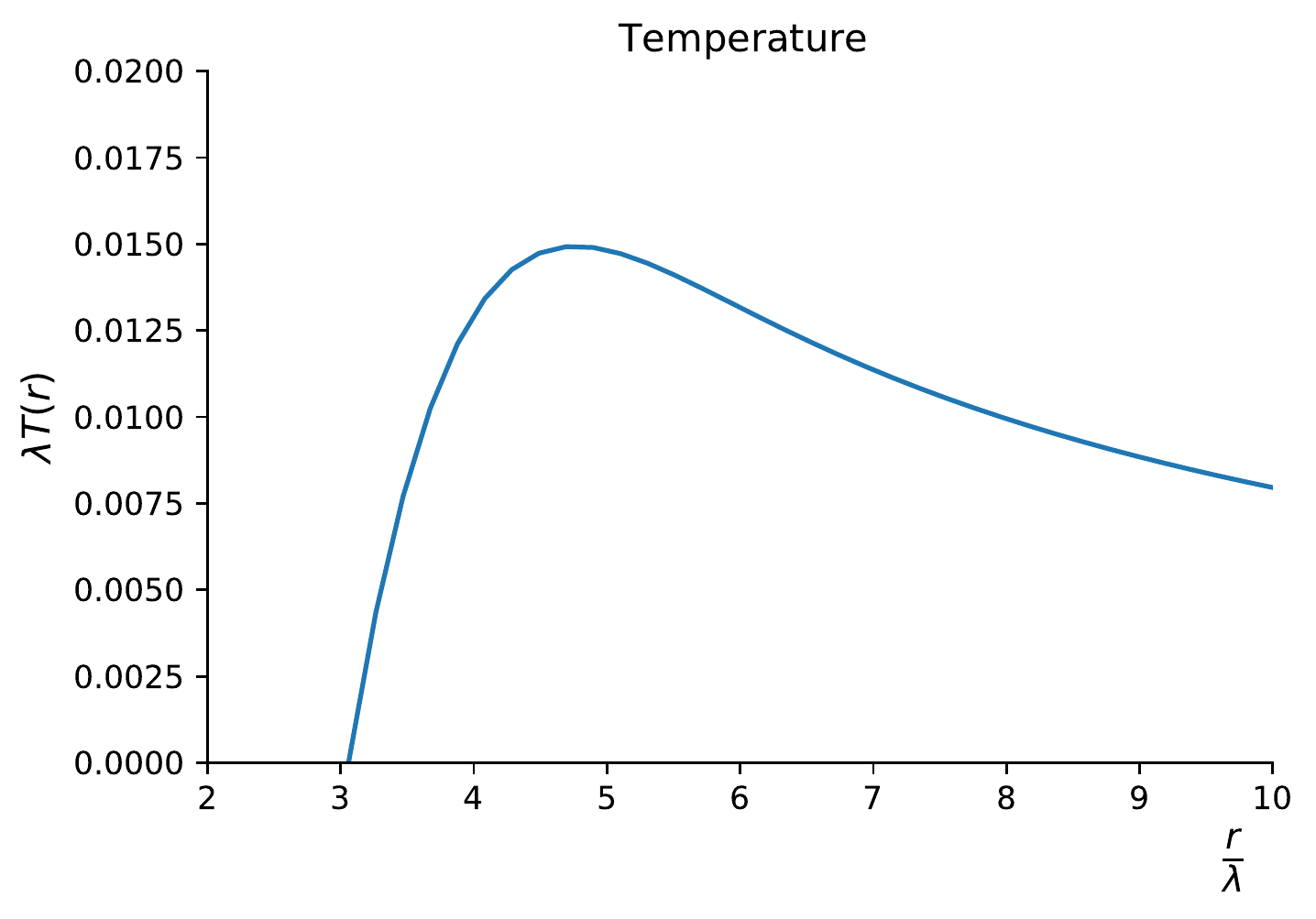}
		\caption{Temperature as a function of the radius, in units of $\lambda$.}
		\label{fig:temperature}
	\end{figure}

	Here three cases are represented, corresponding to zero, one or two intersections with the $r$ axis. These possibilities are discriminated by a minimal mass $M_0$, below which there is no event horizon. 
	It can be shown (see \cite{14}) that if $M<M_0$ there are no naked singularity, despite there being no horizons. The trace of the Einstein equations gives the curvature scalar:
	\begin{equation}
	R(0) = \frac{4 M}{\sqrt{\pi}\lambda^3},
	\end{equation}
	which is nowhere divergent, including at $r=0$. 
	From the equation for $f(r)$ \eqref{f(r)} we can also calculate the Hawking temperature in the usual way:
	\begin{equation}
	T = -\left(\frac{1}{4\pi} \frac{\d g_{00}}{\d r}\right)_{r_H}.
	\end{equation}
	Using the expression \eqref{hor rad} to re-express $M$, we obtain:
	\begin{equation}
	T = \frac{1}{4\pi r_H} \left[1-\frac{r_H^3 e^{-\frac{r_H^2}{4\lambda^2}}}{4\lambda^3 \gamma\left(\frac{3}{2},\frac{r_H^2}{4\lambda^2}\right)}\right]. \label{temp}
	\end{equation}
	The temperature can now be plotted in units of $\lambda$, giving the result in Fig.(\ref{fig:temperature}).
	
	We note that zero temperature corresponds to $r_0$, since it is the minimum of $-g_{00}$ (see figure \ref{fig:horizons}), implying $\d g_{00}/\d r = 0$ in $r_0$. For $r_H < r_0$ there is no horizon and thus the $T$ in Eq.\eqref{temp} cannot be defined. We can see the numerical results in Table \ref{tab:table}, expressed in the Planck units.
	
	\begin{table}[h!]
		\begin{tabular}{lllll}
			$	M_0$ :& $\left(1.6874815942\right.$ & $\pm$ & $\left.10^{-10}\right)$ & $\lambda$ \\
			$	r_{0(H)}$ :& $\left(3.02243\right.$  & $\pm$ & $\left.10^{-5}\right)$  & $\lambda$ \\
			$	r_{0(T)}$ :& $\left(3.0596813\right.$  & $\pm$ & $\left.10^{-7}\right)$  & $\lambda$ \\
			$	T_m$ :& $\left(1.4934245\right.$ & $\pm$ & $\left.10^{-7}\right) 10^{-2}$ & $1/\lambda$ \\
			$	r_m$ :& $\left(4.7663222\right.$ & $\pm$ & $\left.10^{-7}\right)$ & $\lambda$ \\
			$	M_m$ :& $2.41$ &  &  & $\lambda$ \\
		\end{tabular}
		\caption{The values of $r_{0(H)}$ and $M_0$ are calculated numerically from Fig.(\ref{fig:horizons}), while $r_{0(T)}$ is obtained from Fig.(\ref{fig:temperature}). The latter result is different from $r_{0(H)}$, but more precise and convenient to solve for since Eq.\eqref{temp} only involves $r_H$ and $\lambda$, whereas Eq.(\ref{hor rad}) involves also $M$. We suspect this discrepancy is due to the presence of the incomplete gamma function in the numeric (as this does not arise in the models with $\alpha=1$ and $\alpha=3$ discussed in the appendix), but in any case this does not affect our qualitative discussions. The value $M_m$ is obtained analytically from Eq.\eqref{hor rad}.  
			\label{tab:table}}
	\end{table}
	
	Another feature of the peak temperature is the recoil velocity, as analysed in \cite{1} for the $\alpha =1$ model. As explained briefly in Sec.(\ref{intro}), the argument is that a number of quanta is emitted as the black holes evolve from the temperature peak to $T=0$, which spans a relatively narrow range of the mass, thus giving a huge recoil velocity to the black holes. Restoring the physical constants, the number of quanta in SI units is:
	\begin{equation}
	N_q \sim \frac{c^2(M_m-M_0)}{T_m k_B} \approx 48 \lambda^2,
	\end{equation} 
	where $k_B$ is the Boltzmann constant and the masses and temperature used above are the values taken from the Table \ref{tab:table} multiplied by the Planck mass and Planck temperature respectively. Then, we get $N_q  \approx 48$ if we choose $\lambda = 1$ ($L$ is the Planck length $\ell_\text{P}$).
	Calculating the momentum of the emitted quanta as the result of a random walk process, we get the following expression for the recoil velocity of the black hole:
	\begin{flalign}
	v_r &= \frac{M_m-M_0}{M_0 \sqrt{N_q}} c = \frac{\sqrt{\left(M_m-M_0\right)T_m k_B}}{M_0} \notag \\&\approx 1.85 \cdot 10^{7} \frac{1	}{\lambda}~ \text{ms}^{-1}. 
	\end{flalign}
	{The conversion of the masses and temperature is obtained exactly as done before.
		Choosing $L = \ell_\text{P}$} gives $v_r \approx 1.85 \times 10^{4}$ km/s, which is far too large to be a cold dark matter candidate.	
	Indeed, in \cite{dm review}, the circular velocity of dark matter in our galaxy is estimated to be around $218-246$ km/s. If our result is to be comparable with this, then it would require choosing $L \sim 10^2 \ell_\text{P}$.
	
	At this point, one might wonder why we do not consider the speed caused by the recoil as the black hole evaporates from its initial mass down to $M_m$. This is because the speed in this period is negligible. In the model of \cite{1210.6348} for example, it was shown that the speed of a recoiled Schwarzschild black hole is roughly the inverse of the initial mass for most part of the evolution until the last moments of the evaporation. This is due to the fact that the momentum is set by the temperature scale, and as the temperature becomes unbounded towards the end, one can show that the velocity, which is related to the speed via $p=\gamma mv$ in the relativistic regime, will asymptote to the speed of light. In the non-commutative case discussed here, however, the maximum temperature occurs way before Planck temperature, and so the speed is negligible for most part of the evolution before $M_m$ is reached.
	
	Now, according to \cite{14a} and \cite{astro constraint}, if the Hawking temperature of a black hole reaches $T \approx 2.7$ K at the current epoch, it will attain thermal equilibrium with the cosmic microwave background (CMB) photons and thus stop evaporating. While this may be true for larger black holes, it is not what happens in our case when the black holes are microscopic. Indeed, even if the black hole and the universe share the same temperature, they are not actually in thermal equilibrium. This is due to the scattering cross section $\sigma \sim \pi r_0^2  \approx 28$ $\lambda^2$, corresponding to $7.38 \times 10^{-69}$ $\lambda^2$ $\text{m}^2$ in SI units, that is too small for any effective interaction with the CMB radiation. If we want $\sigma$ to be of the order of the Thomson cross section $\sigma_T  \approx 6.6 \times 10^{-29}$ $\text{m}^2$, we would need {$L \sim 10^{20}$} $\ell_\text{P}$, which exceeds even the lower end of the phenomenological constraint $L < 10^{-18}$ m. Therefore, we can conclude that the evaporation of non-commutative black hole would continue as it approaches the asymptotic extremal state $T=0$ and $M=M_0$. Expressing this mass in SI units, it is $M_0 = 3.67 \times 10^{-8}$ {$\lambda$} kg. This value is a little larger than $M_\text{P}$ (the Planck mass) even for $\lambda=1$, so the remnants left by the evaporation, though the same order, are not exactly Planckian.
	
	Now let us see whether primordial black holes (PBHs) could be found around the temperature peak in our present epoch. The mass corresponding to the peak temperature, when the constants are restored, is $M_m = 2.41$ $\lambda M_\text{P}$. In SI units, we have $M_m = 5.25 \times 10^{-8}$ $\lambda$ kg. Following \cite{rice}, black holes with initial masses larger than $10^{15}$ g would stay essentially the same throughout their lifetime and the more massive ones can even be accreted\footnote{While the analysis in \cite{rice} assumes Scharwzschild geometry, due to the smallness of $\lambda$, the behavior of the black holes in non-commutativity spacetimes do not differ substantially from the one in general relativity until the black hole becomes microscopic.}. In fact, the time scale for evaporation is even slightly longer if one takes into account the sparsity\footnote{Sparsity means Hawking radiation is not a continuous emission of particles; instead, most particles slowly ``drip'' out one by one in a random direction, with the average time between successive emissions being larger than the typical time scale set by the energies of the emitted quanta. If radiation were a continuous stream from all directions, there would not be much recoil since the momenta would cancel.} of the radiation \cite{1506.03975}. Thus, this kind of black holes would be too far from the peak for any allowed value of the parameter $\lambda$. 
	
	On the contrary, PBHs with initial masses equal or smaller than the critical mass $M_{cr} = 5.1 \times 10^{14}$ g can evaporate effectively \cite{rice}. If the mass is around $M_{cr}$, the PBH is expected to have evaporated to around $M_0$ by the present epoch (assuming general relativity holds). Therefore, for a quick estimate, taking a slightly higher initial mass means the black hole could be located near the temperature peak if there is a non-commutativity correction. The value of $\lambda$ in its range of validity does not affect substantially the initial mass $M_i \sim M_{cr}$ and so we are free to set $\lambda \sim 10^2$ to match the velocity of dark matter. Black holes with initial masses lower than $M_{cr}$, would be well on their way to become extremal remnants of mass $M_0$.
	
	The bottom line is that the model comes with a non-commutativity scale $\lambda$, the value of which can be inferred from the dark matter velocity if the primordial non-commutative black holes are indeed dark matter particles. We also check the model with $\alpha=1$ and $\alpha=3$ in the Appendix and find the same conclusion holds.

	\section{Conclusion: Microscopic Primordial Black Holes Remain Feasible as Dark Matter Candidate}
	
	In this work we have shown that primordial black holes remain a viable dark matter candidate. The argument in \cite{rice} that microscopic black holes gain too large a velocity towards the end of Hawking evaporation (and therefore cannot behave as cold dark matter) of course depends on the underlying models of quantum gravity phenomenology. Specifically it only applies to those models in which the Hawking temperature of a Schwarzschild-like black hole turns around near the end of the Hawking evaporation, such as the models that employ non-commutative spacetime. Even then, this is true only for \emph{some} values of the non-commutative parameter $\lambda$, and is only relevant if the initial mass of the black hole is small enough -- for otherwise it would not have enough time to evolve past the temperature peak. 
	
	We show that it is possible to choose $\lambda$ such that microscopic black holes that have passed the temperature peak still have velocity that is consistent with dark matter, as required from astrophysical observations. In this simple analysis following the method in \cite{13}, the value of $\lambda$ turns out to be at the order of $\mathcal{O}(10^2)$ (this is also the case for the other non-commutative models explored in the appendix, so it is somewhat robust), which is still within the observational constraint of not having detected a non-commutative spacetime geometry. Interestingly, this is consistent with the argument in \cite{1901.01613} that the scale of non-commutativity could be larger than
	the Planck length.
	
	In a more careful analysis, the time it takes for a black hole to evolve from $M_m$ (the mass corresponding to the temperature peak) to $M_0$ (the final remnant mass) should be taken into account by solving the evolution equation of $\d M/\d t$, though this is not quite feasible in these models since the expression for the horizon cannot be readily obtained. Since the final remnant has zero temperature, the evaporation time from the peak to the final state would likely take an infinite amount of time. In fact, as the temperature drops, the evaporation process becomes slower and slower. This mean that the recoil velocity might be of the form
	\begin{equation}
	v_r \approx \frac{M_m-M_*}{M_* \sqrt{N}}=\frac{\sqrt{(M_m-M_*)T_m}}{M_*},
	\end{equation}
	where $M_* > M_0$ is the current mass of the black hole. 
	We note that $v_r$ is a decreasing function of $M_*$ for fixed $M_m$. Therefore, there is a value near the maximum $M_m$ where it is of the same order of magnitude as the observed velocity, for $\lambda = 1$. For $M_*$ even larger than this value (but still less than $M_m$), this would imply that the recoil velocity could be considerably smaller than expected even if we take the length scale $L$ to be closer to the Planck length.
	
Since astrophysical black holes have angular momentum, and could be charged (not necessarily by ordinary Maxwell field, but conceivably by some hidden U(1) field related to the dark sector), one might wonder whether our conclusions would need to be revise substantially if these effects are included. 
We think that this is unlikely. Charged black holes in the presence of spacetime non-commutativity had been studied by Alavi \cite{0909.1688}. We can repeat our calculation using the charged metric and indeed the qualitative results remain unchanged. From the work of Nicolini and Modesto \cite{1005.5605}  we can also argue that when rotation is included, the result is also qualitatively the same.  See also \cite{Paik}, in which Paik discussed how the angular momentum and charge are lost as we approach the non-commutativity scale. In other words, charge and angular momentum do not affect the conclusion by much  at this scale (they would certainly be important when the black holes are still at the astrophysical scale, and the exact evolution up to near Planck scale would require much detailed modeling).
	
	In any case, we conclude that recoil speed due to Hawking radiation does not rule out primordial microscopic black holes as a dark matter candidate\footnote{After our work appears, another analysis argues that if cosmic expansion is taking into account, the recoil velocity may not be in tension with primordial black hole remnants being cold dark matter \cite{2105.01627}.}.
	Finally we remark that the analysis and discussion in this work is not only applicable to the non-commutative black hole models but also to all other quantum gravitational inspired models as long as the microscopic black holes have a temperature peak and a turnover of the temperature after the peak (whether the end state is some finite temperature state or zero temperature state is not so crucial). One such example is the Schwarzschild-like black hole in asymptotically safe gravity \cite{0602159v1, 1401.4452v1}, which we also briefly discuss in the Appendix. 
	
	\begin{acknowledgments}
		YCO thanks the National Natural Science Foundation of China (No.11922508) for funding support. 
	\end{acknowledgments}
	
	\begin{appendices}
		\section*{Appendix}
		In this Appendix we shall repeat the calculation in Sec.(\ref{main}) for the models in which $\alpha=1$ and $\alpha=3$, respectively. The results for both cases are essentially the same with that of $\alpha=2$ discussed above. Nevertheless it might be useful for future references to have the explicit results shown.

		\subsection{Model with $\alpha=1$}
		An alternative model, obtained by taking $\alpha = 1 $ in the density function \eqref{generic density}, has been investigated in \cite{13} and \cite{1}. The density is given by
		\begin{equation}
		\rho(r) = \frac{M}{(2\lambda)^3\pi} e^{-r/\lambda}.
		\end{equation}
		Proceeding analogously as the $\alpha=2$ model and requiring the metric to be Schwarzschild-like and spherically symmetric, we calculate the $00$-component of the Einstein equations:
		\begin{equation}
		\frac{1+f(r)+f^\prime(r)}{r^2} = 8\pi T_{00} = \frac{M}{\lambda^3} e^{-r/\lambda},
		\end{equation}
		where $g_{00} = f(r) = -g_{rr}^{-1}$ and $T^0_0 = -\rho(r)$.
		The solution of the above equation is:
		\begin{equation}
		f(r) = -1 + \frac{2M}{r}- \frac{M}{r}e^{-\frac{r}{\lambda}}\left(\frac{r^2}{\lambda^2}+\frac{2r}{\lambda}+2\right). \label{f(r)1}
		\end{equation}
		The event horizon is located at $g_{rr} = 0$, i.e., at $f(r)=0$. To find this value, it is again necessary to plot $f(r)$ and compute numerically its intersections with the $r$ axis:
		\begin{figure}[h!]
			\centering
			\includegraphics[scale=0.60]{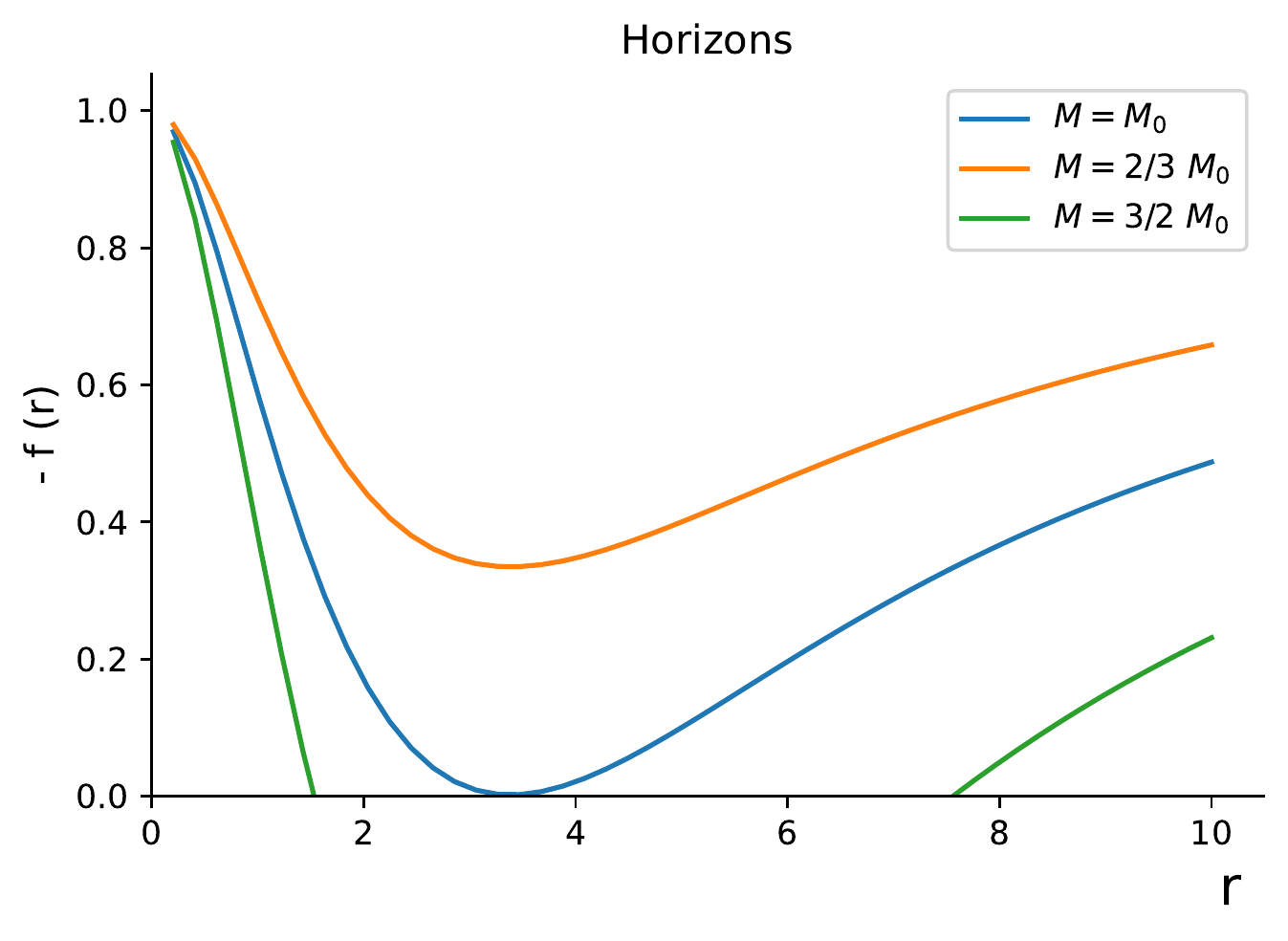}
			\caption{$-f(r)$ as a function of the radius, in units of $\lambda$ for $M=M_0$, $M =2/3$ $M_0$, $M = 3/2$ $ M_0$.}
			\label{fig:horizons1}
		\end{figure}
		\noindent We notice again that there are three cases: zero, one or two intersections with the $r$ axis, corresponding to $M$ less, equal or greater than the minimal mass $M_0$. 
		From Eq.\eqref{f(r)1}, the Hawking temperature is given by
		\begin{equation}
		T = -\left(\frac{1}{4\pi} \frac{\d g_{00}}{\d r}\right)_{r_H}.
		\end{equation}
		From the same equation, Eq.\eqref{f(r)1}, we can also derive an expression for $M(r_H)$ when $f(r)=0$ and substitute it in the temperature equation. The end result is:
		\begin{equation}
		\lambda T = \frac{1}{4\pi x} \frac{x^3+ x^2 + 2x-2\left( e^x-1\right)} {x^2+ 2x-2\left( e^x-1\right)},
		\end{equation}
		where $x = r_H/\lambda$.
		The plot of the temperature, expressed in units of $\lambda$, is given in Fig.(\ref{fig:temperature1}) below.
		\begin{figure}[h!]
			\centering
			\includegraphics[scale=0.60]{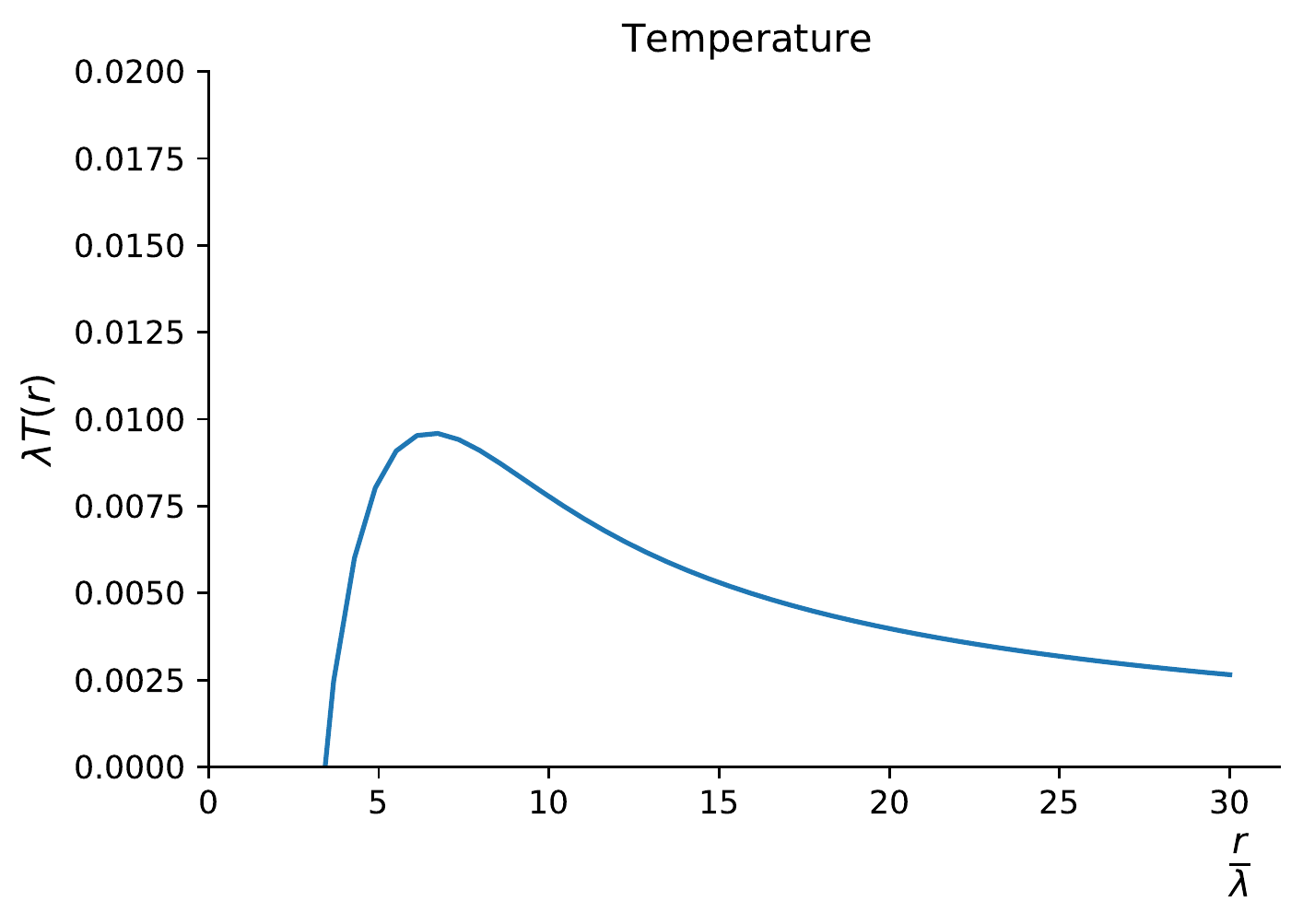}
			\caption{Temperature as a function of the radius, in units of $\lambda$.}
			\label{fig:temperature1}
		\end{figure}
		This plot is very similar to Fig.(\ref{fig:temperature}) in the previous model with $\alpha=2$, with the absolute zero of the temperature corresponding to $r_0$ and no horizons present for $r_H < r_0$, as we would expect. The numerical results in the Planck units for this model are collected in Table \ref{tab:table1}.\\
		\begin{table}[h!]
			\begin{tabular}{lllll}
				$	M_0$ :& $\left(2.5747013735\right.$ & $\pm$ & $\left.10^{-10}\right)$ & $\lambda$ \\
				$	r_{0(H)}$ :& $\left(3.38363\right.$  & $\pm$ & $\left.10^{-5}\right)$  & $\lambda$ \\
				$	r_{0(T)}$ :& $\left(3.3836342\right.$  & $\pm$ & $\left.10^{-7}\right)$  & $\lambda$ \\
				$	T_m$ :& $\left(9.6023522\right.$ & $\pm$ & $\left.10^{-7}\right) 10^{-3}$ & $1/\lambda$ \\
				$	r_m$ :& $\left(6.5442890\right.$ & $\pm$ & $\left.10^{-7}\right)$ & $\lambda$ \\
				$	M_m$ :& $2.72$ &  &  & $\lambda$ \\
			\end{tabular}
			
			\caption{The values of $r_{0(H)}$ and $M_0$ are calculated numerically from Fig.(\ref{fig:horizons1}), while $r_{0(T)}$ is obtained from Fig.(\ref{fig:temperature1}). Contrary to the results in Table \ref{tab:table}, $r_{0(H)}$ and $r_{0(T)}$ are compatible. The value $M_m$ is obtained analytically.  \label{tab:table1}}
		\end{table}

		Next we analyze the recoil velocity of the primordial black holes due to Hawking radiation. The number of emitted quanta from the peak temperature to $T = 0 $ is:
			\begin{equation}
			N_q \sim \frac{c^2(M_m-M_0)}{T_m k_B}  \approx 87 \lambda^2,
			\end{equation} 
			where the quantities in Table \ref{tab:table1} are expressed in SI units.
		The recoil velocity the black hole due to the emission of these quanta is
			\begin{flalign}
			v_r &= \frac{M_m-M_0}{M_0 \sqrt{N_q}} c = \frac{\sqrt{\left(M_m-M_0\right)T_m k_B}}{M_0} \notag \\ &\approx 1.04 \times 10^{7} \frac{1	}{\lambda} \text{ms}^{-1}.
			\end{flalign}
			Choosing $\lambda = 1 $ gives $v_r \approx 1.04 \times 10^{4}$ km/s, then to match the results for the dark matter velocity as given in \cite{dm review}, we again need $\lambda \sim 10^2$.
		
		We can draw the same conclusions as in the $\alpha =2$ model, namely primordial black holes with initial masses around the critical mass can be found near the peak of the temperature for every allowed value of $\lambda$ and they subsequently evaporate to a remnant with temperature $T=0$ and mass $M = M_0 = 5.60\times 10^{-8}$ $\lambda$ kg (a value larger than the Planck mass even for $L = \ell_\text{P}$). Also, primordial black holes with higher and lower masses are found far from the peak or as remnant near $T=0$, respectively.
		
		\subsection{Model with $\alpha=3$}
\begin{figure}[h!]
			\centering
			\includegraphics[scale=0.60]{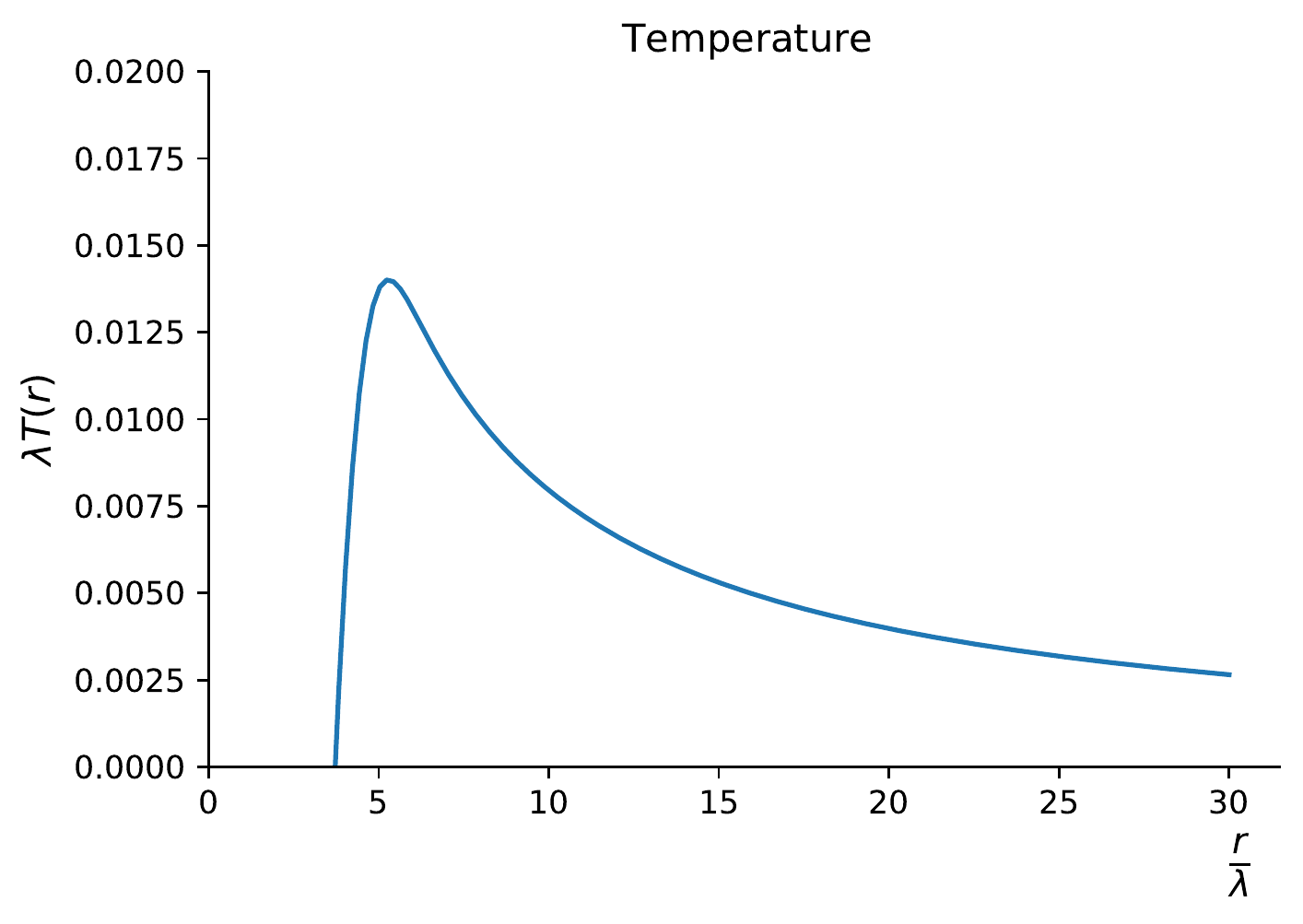}
			\caption{Temperature as a function of the radius, in units of $\lambda$.}
			\label{fig:temperature3}
		\end{figure}

		Another model results from taking $\alpha = 3 $ in the density function \eqref{generic density} and was proposed in \cite{8}.
		The density was written as
		\begin{equation}
		\rho(r) = \epsilon_0 e^{-\left(r/r_*\right)^3},
		\end{equation}
		where
		\begin{equation}
		\epsilon_0 = \frac{3 c^2 M}{4\pi r_*^3} \equiv \frac{M}{36\pi \lambda^3},
		\end{equation}
		in which $M$ is a constant that indicates the total mass if this was a Schwarzschild black hole. In the above expression, we renamed the theory's parameter $\epsilon_0$ so that we have the same structure as the other models. Also, we took $c=1$, consistently with our choice of units. The density then takes the following form:
		\begin{equation}
		\rho(r) = \frac{M}{36\pi \lambda^3} e^{-\left(r/(3\lambda)\right)^3}.
		\end{equation}
		The Einstein equations give
		\begin{equation}
		\frac{1+f(r)+f^\prime(r)}{r^2} = 8\pi T_{00} = \frac{2M}{9 \lambda^3} e^{-\left(r/(3\lambda)\right)^3},
		\end{equation}
		where, again, $g_{00} = f(r) = -g_{rr}^{-1}$ and $T^0_0 = -\rho(r)$.
		The solution is:
		\begin{equation}
		f(r) = -1 + \frac{2M}{r}\left[1-e^{-\left(r/(3\lambda)\right)^3}\right] \label{f(r)3}.
		\end{equation}
		The event horizon is once again found by imposing $f(r)=0$. To obtain it numerically, we plot $f(r)$ and observe its intersections with the $r$ axis:
		\begin{figure}[h!]
			\centering
			\includegraphics[scale=0.60]{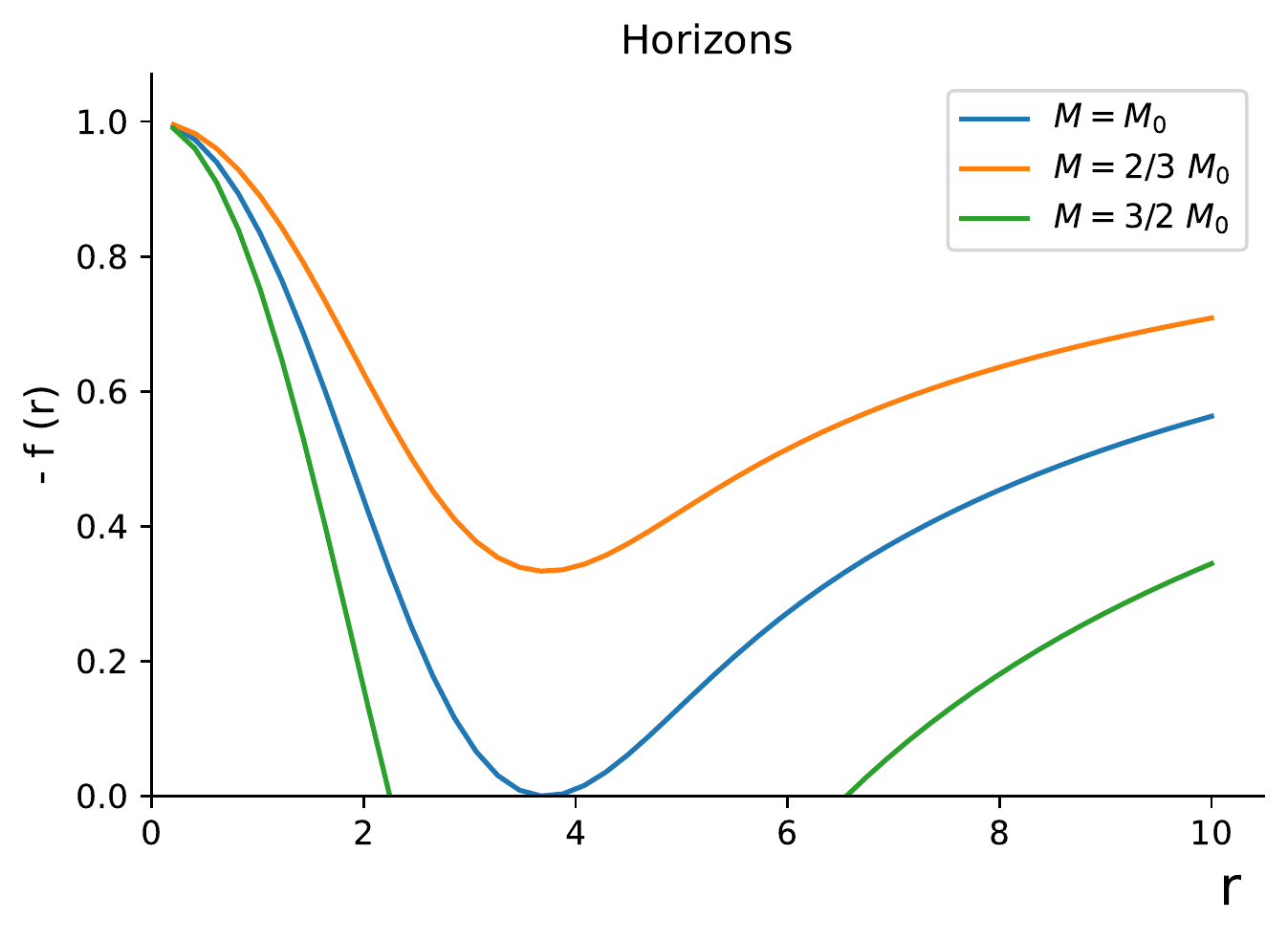}
			\caption{$-f(r)$ as a function of the radius, in units of $\lambda$ for $M=M_0$, $M =2/3$ $M_0$, $M = 3/2$ $ M_0$.}
			\label{fig:horizons3}
		\end{figure}
		\noindent Horizons are formed only for $M\geqslant M_0$, and for $M= M_0$ they coincide. Next, the temperature can be calculated using $f(r)$ in Eq.(\ref{f(r)3}):
		\begin{equation}
		T = -\left(\frac{1}{4\pi} \frac{\d g_{00}}{\d r}\right)_{r_H}.
		\end{equation}
		Substituting the expression for $M(r_H)$ easily obtainable from $f(r)=0$, we have:
		\begin{equation}
		T = \frac{1}{4\pi r_H} \left[1-\frac{3}{e^{\left(r_H/(3\lambda)\right)^3}-1}\left(\frac{r_H}{3\lambda}\right)^3\right].
		\end{equation}
		The plot of the temperature, expressed in units of $\lambda$, is given in Fig.(\ref{fig:temperature3}).
		
		The plot is analogous to the models studied previously (in figures \ref{fig:temperature} and \ref{fig:temperature1}), with the absolute zero of the temperature corresponds to $r_0$, and there is no horizon for $r_H < r_0$. The numerical results in the Planck units for this model are listed in Table \ref{tab:table3} below.\\
		\begin{table}[h!]
			\begin{tabular}{lllll}
				$	M_0$ :& $\left(2.1845877235\right.$ & $\pm$ & $\left.10^{-10}\right)$ & $\lambda$ \\
				$	r_{0(H)}$ :& $\left(3.71817\right.$  & $\pm$ & $\left.10^{-5}\right)$  & $\lambda$ \\
				$	r_{0(T)}$ :& $\left(3.7181714\right.$  & $\pm$ & $\left.10^{-7}\right)$  & $\lambda$ \\
				$	T_m$ :& $\left(1.4009975\right.$ & $\pm$ & $\left.10^{-7}\right) 10^{-2}$ & $1/\lambda$ \\
				$	r_m$ :& $\left(5.2883955\right.$ & $\pm$ & $\left.10^{-7}\right)$ & $\lambda$ \\
				$	M_m$ :& $2.66$ &  &  & $\lambda$ \\
			\end{tabular}
			\caption{The values of $r_{0(H)}$ and $M_0$ are calculated numerically from Fig.(\ref{fig:horizons3}), while $r_{0(T)}$ is obtained from Fig.(\ref{fig:temperature3}). Contrary to the results in Table \ref{tab:table}, and similar to that in Table \ref{tab:table1}, $r_{0(H)}$ and $r_{0(T)}$ are compatible. The value $M_m$ is obtained analytically.  \label{tab:table3}}
		\end{table}
		
		The recoil velocity is once again calculated from the number of emitted quanta from the peak temperature to $T = 0$:
			\begin{equation}
			N_q \sim \frac{c^2(M_m-M_0)}{T_m k_B} \approx 34 \lambda^2,
			\end{equation} 
		in SI units.
		The recoil velocity is then:
			\begin{flalign}
			v_r &= \frac{M_m-M_0}{M_0 \sqrt{N_q}} c = \frac{\sqrt{\left(M_m-M_0\right)T_m k_B}}{M_0} \notag \\ &\approx 1.12 \times 10^{7} \frac{1	}{\lambda} \text{ms}^{-1}.
			\end{flalign}
			Choosing $\lambda = 1$ gives $v_r \approx 1.12 \times 10^{4}$ km/s, then to match the results in \cite{dm review} we need $\lambda \sim 10^2$.
		
		The model is again overall very similar to the previous two discussed.
		
		\subsection{Model in Asymptotically Safe Gravity}
		
\begin{figure}[h!]
			\centering
			\includegraphics[scale=0.60]{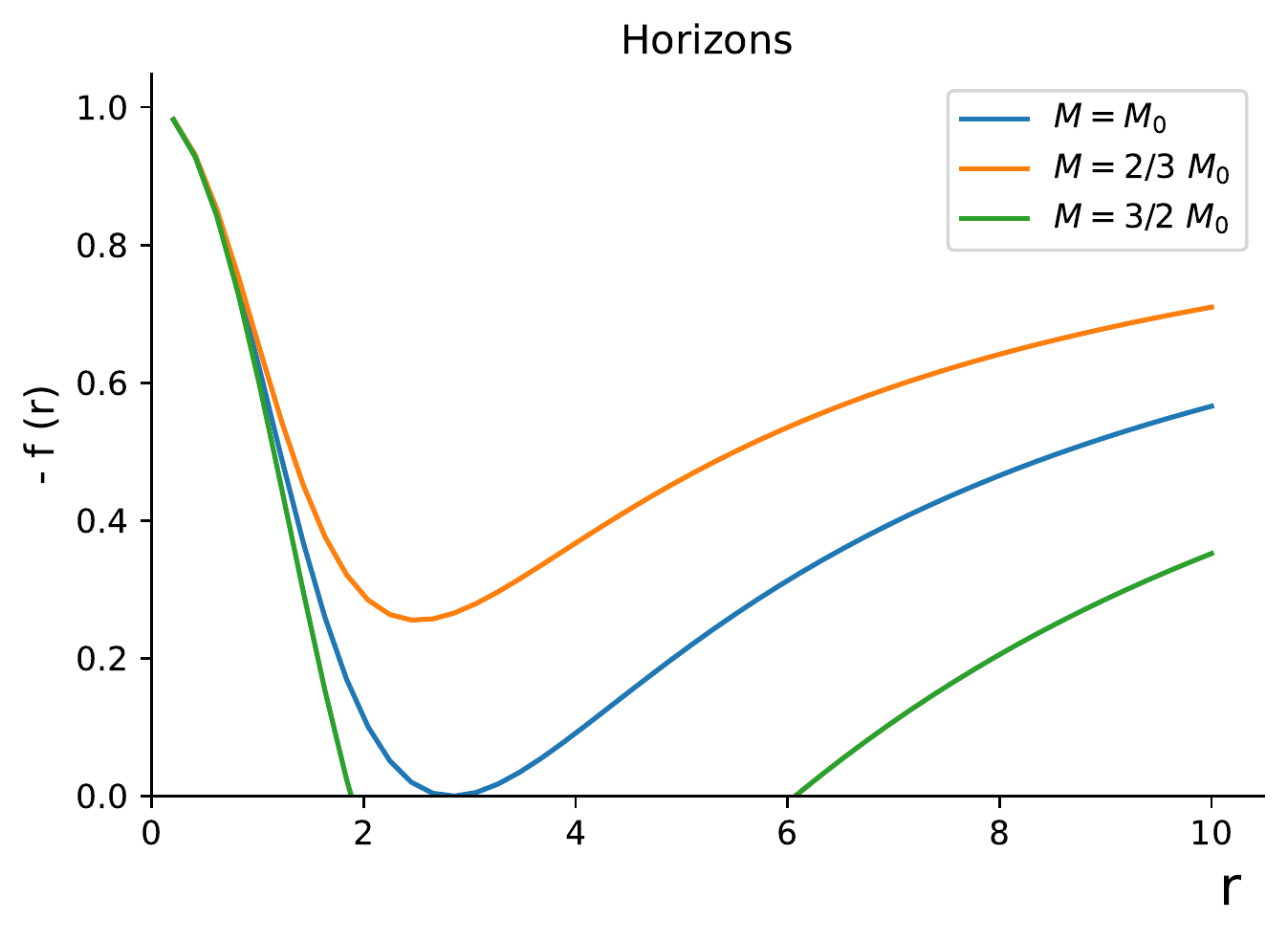}
			\caption{$-f(r)$ as a function of the radius, in natural units and for $\omega=1$, corresponding to the values $M=M_0$, $M =2/3$ $M_0$, $M = 3/2$ $ M_0$.}
			\label{fig:horizons asy}
		\end{figure}

\begin{figure}[h!]
			\centering
			\includegraphics[scale=0.60]{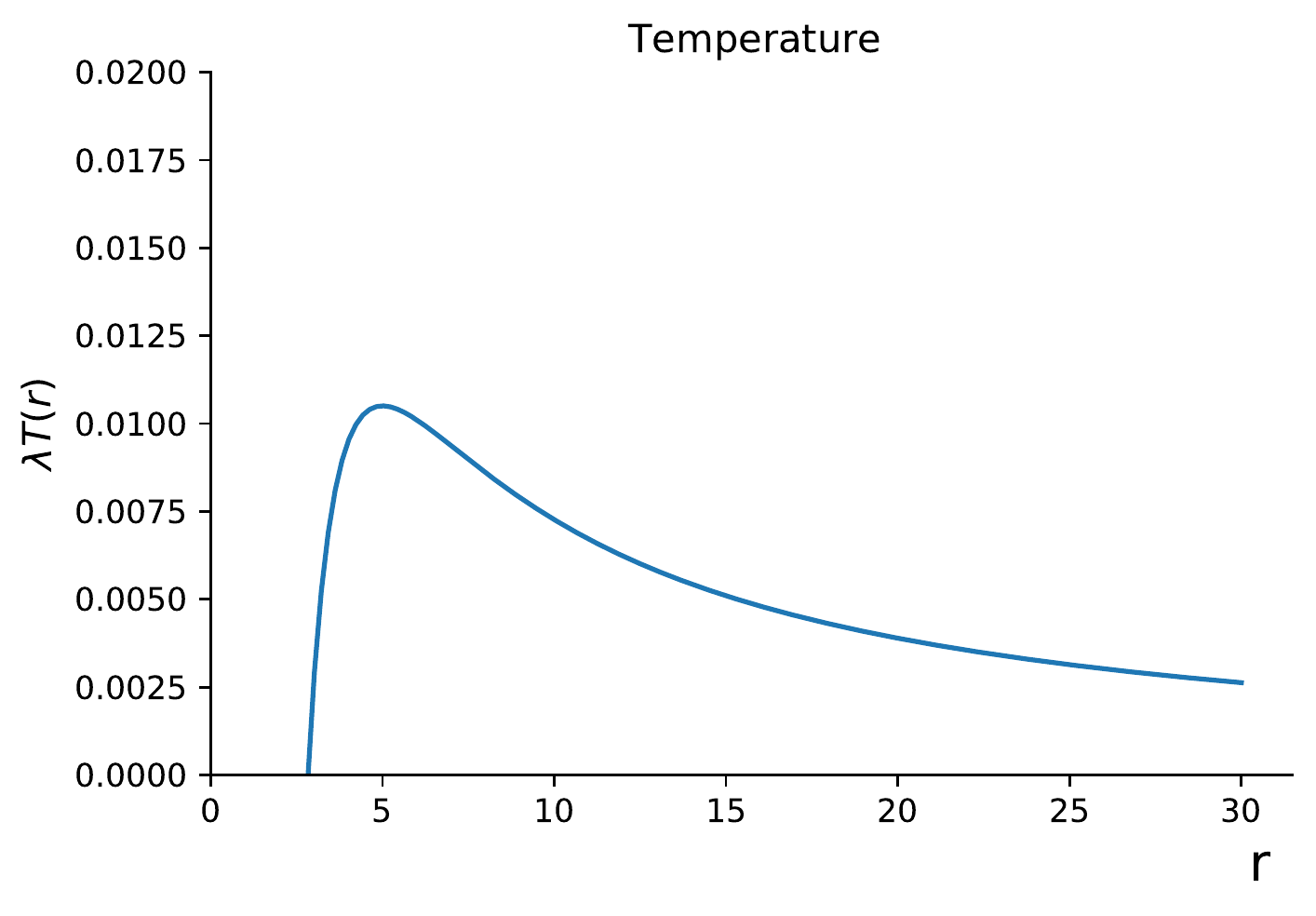}
			\caption{Temperature as a function of the radius, in natural units with $\omega=1$.}
			\label{fig:temperature asy}
		\end{figure}

		Black hole properties arising from asymptotic safety have been analysed starting with the article by Bonanno and Reuter in \cite{0602159v1}. In this treatment, we follow \cite{1401.4452v1}. The radial function $f(r)$ under RG-improvement of a Schwarzschild black hole is:
		\begin{equation}
		f(r) = -1 + \frac{4MG r^2}{2r^3 + \omega G\left(2r+9GM\right)}, \label{f(r) asy}
		\end{equation}
		where $\omega$ is a parameter of the theory and $G$ is the usual Newton's constant, which can be set to $1$ to use natural units.
		Imposing $f(r)=0$ gives the position of the event horizon, obtained numerically from the plot in Fig.(\ref{fig:horizons asy}), and the mass $M_0$:
		\begin{equation}
		M_0 = 2r \frac{r^2+\omega}{4r^2-9 \omega}. \label{M_0 asy}
		\end{equation}
		Horizons are formed only for $M\geqslant M_0$, and for $M= M_0$ they coincide. Next, the temperature can be calculated using $f(r)$ in \eqref{f(r) asy}:
		\begin{equation}
		T = -\frac{1}{4\pi} \left(\frac{\d f(r)}{\d r}\right)_{r_H},
		\end{equation}
		giving the result:
		\begin{equation}
		T = \frac{Mr}{2\pi} \frac{r^3-\omega\left(r+9M\right)}{\left[r^3+\omega\left(r-\frac{9}{2} M\right)\right]^2},
		\end{equation}
		where one could also substitute the expression for $M_0$ in \eqref{M_0 asy}.
		The plot of the temperature, expressed in natural units with $\omega=1$, is given in Fig.(\ref{fig:temperature asy}).

		The plot is analogous to those of the non-commutativity models, with the absolute zero of the temperature corresponding to $r_0$, and the absence of horizons for $r_H < r_0$. The numerical results with $\omega=1$ are listed in Table \ref{tab:table asy} below.\\
		\begin{table}[h!]
			\begin{tabular}{lllll}
				$	M_0$ :& $\left(2.2135416027\right.$ & $\pm$ & $\left.10^{-10}\right)$ & $\omega m_\text{p}$ \\
				$	r_{0(H)}$ :& $\left(2.8338\right.$  & $\pm$ & $\left.10^{-4}\right)$  & $\omega \ell_\text{p}$ \\
				$	r_{0(T)}$ :& $\left(2.8337594\right.$  & $\pm$ & $\left.10^{-7}\right)$  & $\omega \ell_\text{p}$ \\
				$	T_m$ :& $\left(1.0504229\right.$ & $\pm$ & $\left.10^{-7}\right) 10^{-2}$ & $T_\text{p}$ \\
				$	r_m$ :& $\left(5.0020168\right.$ & $\pm$ & $\left.10^{-7}\right)$ & $\omega \ell_\text{p}$ \\
				$	M_m$ :& $2.86$ &  &  & $\omega m_\text{p}$ 
			\end{tabular}
			\caption{The values of $r_{0(H)}$ and $M_0$ are calculated numerically from figure \ref{fig:horizons asy}, while $r_{0(T)}$ is obtained from figure \ref{fig:temperature asy}. $r_{0(H)}$ and $r_{0(T)}$ are compatible. The value $M_m$ is obtained analytically.  \label{tab:table asy}}
		\end{table}

		The recoil velocity the black hole gets from the emission of quanta in the last part of its evolution then gives:
		\begin{equation}
		N_q \sim \frac{c^2(M_m-M_0)}{T_m k_B} \approx 61,
		\end{equation} 
		when $\omega = 1$. The recoil velocity is, under the same assumptions:
		\begin{flalign}
		v_r &= \frac{M_m-M_0}{M_0 \sqrt{N_q}} c = \frac{\sqrt{\left(M_m-M_0\right)T_m k_B}}{M_0} \notag \\ &\approx 1.11 \times 10^{4} \text{ }\frac{km}{s}.
		\end{flalign}
		Choosing $\omega = 4000$ gives $v_r \approx 218$ km/s.
		
	\end{appendices}

\end{document}